\documentclass[10pt,twocolumn]{article}
\usepackage[letterpaper,margin=0.75in]{geometry} 
\usepackage{graphicx}
\usepackage{amsmath}
\usepackage{url}
\usepackage[english]{babel}
\usepackage[utf8]{inputenc}
\usepackage[T1]{fontenc}
\usepackage{hyperref}
\usepackage{xcolor}
\usepackage{caption}
\usepackage{csquotes}
\usepackage{siunitx}
\usepackage{float}
\usepackage{booktabs}
\usepackage[export]{adjustbox}
\usepackage{bookmark}
\usepackage{multirow}
\usepackage{caption}
\usepackage{orcidlink}

\title{Performance and Storage Analysis of CRYSTALS‑Kyber (ML-KEM) as a Post‑Quantum Replacement for RSA and ECC}
\author{
  \textbf{Nicolas \textsc{Rodriguez-Alvarez}}\orcidlink{0009-0002-6804-386X}\\[2pt]
  \small IES Parquesol \\
  \small \href{mailto:nicolas.rodalv@educa.jcyl.es}{nicolas.rodalv@educa.jcyl.es}
}

\date{} 

\newcommand{\perfstyle}{
  \footnotesize
  \setlength{\tabcolsep}{1.5pt}
}

\begin{document}

\twocolumn[
  \maketitle
  \begin{abstract}
      The steady advancement in quantum computer error correction technology has pushed the current record to 48 stable logical qubits, bringing us a little closer to machines capable of running Shor’s algorithm at scales that threaten RSA and ECC cryptography. While the timeline for developing such quantum computers remains uncertain, the cryptographic community must prepare for the transition to quantum-resistant algorithms. CRYSTALS-Kyber, standardized by NIST in 2022, represents a leading post-quantum cryptographic solution, but widespread adoption faces significant challenges. If this migration follows patterns similar to the SHA-1 to SHA-2 transition, organizations may experience prolonged periods of vulnerability, with substantial security and economic consequences. This study evaluates Kyber’s practical viability through performance testing across various implementation schemes, utilizing only standard built-in processor acceleration features (AES-NI, ASIMD) without any specialized hardware additions. Our findings demonstrate that Kyber provides robust security guarantees against quantum attacks while maintaining acceptable performance profiles for most contemporary applications, utilizing only commodity hardware with manufacturer-provided acceleration capabilities.
  \end{abstract}
  \vspace{1em}
]

\section{Introduction}
The theoretical foundations established by Shor and Grover \cite{365700,10.1145/237814.237866} have evolved from academic concepts to practical concerns as quantum computing hardware continues to advance. While experts debate the timeline for achieving fault-tolerant quantum computers capable of running Shor’s algorithm at scale, the cryptographic community faces an urgent imperative: the transition to quantum-resistant algorithms cannot wait for quantum computers to become operational \cite{gri2024_quantum_timeline,gidney2025factor2048bitrsa}.
Historical precedents in cryptographic transitions offer sobering lessons about the challenges ahead. The migration from SHA-1 to SHA-2, initiated in 2005 following the discovery of collision vulnerabilities \cite{10.1007/11535218_2,cryptoeprint:2017/190,nist:fips180-4}, took over a decade to complete \cite{palmer_sleevi_2014}, with many organizations maintaining vulnerable systems well beyond recommended timelines \cite{palmer_sleevi_2014}. This prolonged transition period exposed numerous systems to security risks and highlighted the substantial economic and operational costs associated with delayed cryptographic upgrades. If the transition to post-quantum cryptography follows similar patterns, organizations may face extended periods of vulnerability to quantum attacks, with potentially catastrophic consequences for digital infrastructure, financial systems, and national security.
In response to this challenge, the National Institute of Standards and Technology (NIST) initiated a comprehensive standardization process for post-quantum cryptographic algorithms, culminating in the selection of CRYSTALS-Kyber as the primary Key Encapsulation Mechanism (KEM) standard in 2022 \cite{nist2024fips203}. Unlike RSA and ECC, which derive their security from number-theoretic problems vulnerable to quantum attacks, Kyber’s security relies on the learning with errors (LWE) problem over module lattices—a mathematical foundation believed to be resistant to both classical and quantum computational attacks \cite{8406610}.
However, theoretical quantum resistance alone does not guarantee practical adoption. The success of any cryptographic standard depends critically on its performance characteristics, storage requirements, and compatibility with existing hardware infrastructure. Previous post-quantum proposals have faced significant barriers to adoption due to excessive computational overhead, prohibitive key sizes, or requirements for specialized hardware acceleration. To address these concerns and evaluate Kyber’s practical viability as a replacement for current cryptographic standards, this study conducts a comprehensive performance analysis across multiple architectures and implementation scenarios.
This research contributes to the post-quantum cryptography transition by providing empirical evidence of Kyber’s performance characteristics under realistic deployment conditions. Our evaluation methodology utilizes standard hardware acceleration features available in commodity processors (AES-NI, AVX2, ASIMD) without requiring specialized, quantum-resistant hardware additions, ensuring that our findings accurately reflect the performance organizations can expect during real-world deployment. By comparing Kyber’s performance against equivalent-security implementations of RSA-7680 and SECP384R1 (ECC) across both x86\_64 and ARM64 architectures, we establish benchmarks that inform migration planning and risk assessment for organizations preparing for the post-quantum era.

\section{Background}
\subsection{Classical Cryptographic Schemes}
Modern public-key cryptography relies on two primary mathematical foundations that are vulnerable to quantum attacks. \textbf{RSA} (Rivest-Shamir-Adleman) cryptography derives its security from the computational intractability of the integer factorization problem—the difficulty of decomposing large composite numbers into their constituent prime factors. This problem becomes exponentially harder as key sizes increase, making RSA-2048 and RSA-4096 computationally infeasible to break with classical computers within reasonable timeframes \cite{Lenstra2001,Rivest1978-nz}.
\textbf{Elliptic Curve Cryptography} (ECC) offers an alternative approach based on the elliptic curve discrete logarithm problem (ECDLP) \cite{Miller2007-pn}. ECC achieves equivalent security levels to RSA with significantly smaller key sizes—a 256-bit ECC key provides comparable security to a 3072-bit RSA key. The SECP384R1 curve, standardized by the Standards for Efficient Cryptography (SEC), represents a widely deployed ECC implementation offering 192-bit security strength according to NIST guidelines \cite{nsa-suite-b}.
Both RSA and ECC implementations typically employ hybrid encryption schemes that combine asymmetric and symmetric cryptography. In these systems, a Key Encapsulation Mechanism (KEM) securely exchanges a symmetric key using public-key methods. At the same time, a Data Encapsulation Mechanism (DEM) handles bulk data encryption using faster symmetric algorithms, such as AES or ChaCha20, for which no practical attack is known.

\subsection{Quantum Algorithms}
Shor’s algorithm, formulated by Peter Shor in 1994, marked a milestone in quantum computing theory by showing that a sufficiently large, error-free quantum computer could factor large integers efficiently. Whereas the best-known classical algorithms run in subexponential time, Shor’s algorithm runs in “polynomial” time, approximately $O(\log(N)^3)$ depending on implementation details, making the cryptographic keys based on large‑N factorization (RSA) and on the elliptic curve discrete logarithm problem (ECC) effectively breakable in negligible time \cite{Shor_1997}. Meanwhile, Grover’s algorithm, introduced by Lov Grover in 1996, provides a quadratic speedup for unstructured search, reducing the classical cost of $O(N)$ to $O(\sqrt{N})$. Although not initially intended for factorization, Grover’s amplitude‑amplification technique can be used to optimize specific subroutines within Shor’s method or to accelerate searches among partial solutions generated by Shor’s quantum circuit \cite{10.1145/237814.237866}. In theory, combining Shor’s and Grover’s algorithms could optimize the number of iterations and resource usage.

\subsection{“Standard” hardware accelerations}
In this study, the default hardware accelerations provided by the CPU were utilized. The following section provides a comprehensive explanation of these accelerations.
\subsubsection{Intel}
Intel\textregistered\ Advanced Encryption Standard Instructions (\textbf{AES-NI}): This hardware acceleration provides a speedup of 3 to 10x over an entirely software implementation using AES \cite{intel2012aes}.

Intel\textregistered\ Advanced Vector Extensions (\textbf{AVX/AVX2}): Intel’s vector instruction set for SIMD vector operations \cite{intel21sdmv2}. The Kyber implementation used in this study leverages AVX2 instructions to accelerate its core lattice-based computations, resulting in significant performance improvements.

Intel\textregistered\ Secure Key (\textbf{RDRAND/RDSEED}): On‑chip, NIST‑certified random number generator instructions for a high‑quality entropy source in key generation \cite{intel21sdmv2}.

Intel\textregistered\ Carry‑Less Multiplication (\textbf{PCLMULQDQ}): Provides a single-cycle, hardware‑accelerated carry‑less 64x64‑bit multiply, used in GCM and other Galois‑field operations \cite{intel21sdmv2}.

\subsubsection{ARM}
\textbf{AES}: ARM v8 Cryptography Extensions add AESE/AESD/AESMC/AESIMC instructions for single‑round encryption/decryption and key‑schedule support \cite{arm2014ddi0501f}.

\textbf{SHA1/SHA2}: SHA1C / SHA1P / SHA1M / SHA256H / SHA256SU instructions accelerate SHA‑1 and SHA‑224/256 hashing \cite{arm2014ddi0501f}.

Polynomial Multiply (\textbf{PMULL}): 64x64‑bit carry‑less multiply for efficient GCM‑mode Galois‑field operations \cite{arm2014ddi0501f}.

Advanced SIMD (\textbf{ASIMD}): The “NEON” AArch64 SIMD unit for 128‑bit vector arithmetic, logical, and data‑rearrangement operations \cite{arm2014ddi0501f}.

Half‑Precision SIMD (\textbf{ASIMDHP}): Extension enabling SIMD operations on 16‑bit floating‑point data types \cite{arm2014ddi0501f}.

\section{Cryptographic Schemes}
To ensure a fair comparison, the following widely known and commonly used algorithms will be employed: RSA-7680, ECDH-SECP384R1 (ECC), and ML-KEM768 (Kyber). They all have the same security bit strengths according to the NIST \cite{nsa-suite-b}.

For a practical and fair performance evaluation, it is crucial to test these algorithms as they would be used in a real-world application. Asymmetric cryptography is typically used not for bulk data encryption, but to secure a shared symmetric key or shared secret. This is known as hybrid encryption \cite{cryptoeprint:2001/108}; it was implemented according to the NIST recommendations \cite{Barker2020-dj}.

Therefore, this study evaluates each algorithm within a \textbf{hybrid encryption scheme}, which combines a Key Encapsulation Mechanism (KEM) for the asymmetric
part, and a Data Encapsulation Mechanism (DEM) for the symmetric part.

To ensure consistency, the same DEM —\textbf{ChaCha20-Poly1305} authenticated encryption with associated data (AEAD) cipher \cite{rfc8439}— was used for all three schemes.

A key differentiator between these schemes lies in the mathematical problems that underpin their security. RSA’s security relies on the presumed difficulty of the \textbf{integer factorization problem} \cite{Rivest1978-nz}, while ECC’s is based on the \textbf{elliptic curve discrete logarithm problem (ECDLP)} \cite{Koblitz1987-ii,Miller2007-pn}. Both of these problems are known to be efficiently solvable by a sufficiently large quantum computer using Shor’s algorithm \cite{Shor_1997}.

In contrast, CRYSTALS-Kyber’s security is based on the hardness of solving the \textbf{learning with errors (LWE) problem} over module lattices. The LWE problem is widely believed to be resistant to attacks from both classical and quantum computers, which forms the foundation of its post-quantum security claims \cite{nist2024fips203}.

\section{Benchmarking Methodology}
This section outlines the environment and procedures used for evaluating the performance of the cryptographic schemes.
\subsection{Hardware}
To ensure the veracity of the results, the two most popular architectures were tested on two separate systems: one using the ARM64 architecture—commonly found in portable devices such as the iPhone, Apple Silicon Macs, and the Raspberry Pi—and the other using the x86\_64 architecture, which is prevalent in servers and desktop computers built with Intel\textregistered\ or AMD\textregistered\ processors.
This comparison does not consider hardware-accelerated implementations of certain cryptographic operations, such as RSA ones, which can significantly improve performance \cite{Shantilal2005}. The standard ones that the CPU manufacturer implements have been enabled.
\begin{table}[H]
\centering
\caption{Key Hardware Specifications of x86\_64 Test System}
\label{tab:x86_hardware}
\begin{tabular}{ll}
\toprule
\textbf{Feature} & \textbf{Specification} \\
\midrule
Architecture & x86\_64 (64-bit mode) \\
CPU Vendor & Intel\textregistered\ \\
Processor Model & Xeon E5-2686 v4 \\
Base Frequency & 2.30 GHz \\
CPU Cores & 1 \\
Threads per Core & 1 \\
\\
\multicolumn{2}{l}{\textbf{Cache Hierarchy}} \\
L1 Data Cache & 32 KiB \\
L1 Instruction Cache & 32 KiB \\
L2 Cache & 256 KiB \\
L3 Cache & 45 MiB \\
\\
\multicolumn{2}{l}{\textbf{Cryptography-Relevant Instruction Sets}} \\
AES-NI & Supported \\
AVX/AVX2 & Supported \\
PCLMULQDQ & Supported \\
RDRAND & Supported \\
\bottomrule
\end{tabular}

\end{table}
\begin{table}[H]
\centering
\caption{Key Hardware Specifications of ARM64 Test System}
\label{tab:arm_hardware}
\begin{tabular}{ll}
\toprule
\textbf{Feature} & \textbf{Specification} \\
\midrule
Architecture & ARM aarch64 (64-bit) \\
CPU Vendor & ARM \\
Processor Model & Neoverse-N1 \\
CPU Cores & 2 \\
Threads per Core & 1 \\
Stepping & r3p1 \\
\\
\multicolumn{2}{l}{\textbf{Cache Hierarchy}} \\
L1 Data Cache & 128 KiB (64 KiB/core) \\
L1 Instruction Cache & 128 KiB (64 KiB/core) \\
L2 Cache & 2 MiB (1 MiB/core) \\
L3 Cache & 32 MiB (shared) \\
\\
\multicolumn{2}{l}{\textbf{Cryptography-Relevant Instruction Sets}} \\
AES & Supported \\
SHA1/SHA2 & Supported \\
PMULL (Polynomial Multiply) & Supported \\
ASIMD (Advanced SIMD) & Supported \\
ASIMDHP (FP16 support) & Supported \\
\bottomrule
\end{tabular}
\end{table}

The benchmarking environment employs two distinct systems representing prevalent computing architectures: an x86\_64 platform (Intel\textregistered\ Xeon E5-2686 v4) and an ARM64 platform (ARM Neoverse-N1). Both configurations deliberately emulate general-purpose CPU scenarios by disabling hardware acceleration for asymmetric cryptographic operations (e.g., RSA modular exponentiation units), ensuring fair algorithmic comparisons under standardized software implementations. The x86\_64 system reflects server/desktop profiles with a single-core setup (2.3 GHz base frequency) and a large 45 MiB shared L3 cache, while the ARM64 system mirrors edge/IoT constraints with a dual-core design and 32 MiB shared L3 cache. Crucially, both platforms support modern cryptographic instruction sets: x86\_64 leverages AES-NI, AVX/AVX2, and PCLMULQDQ, whereas ARM64 utilizes AES, PMULL, and ASIMD extensions. For these benchmarks, all available accelerations, including AVX2 for the Kyber implementation, were used where applicable by the cryptographic libraries. These instructions were fully enabled during testing to reflect real-world deployment conditions. The absence of \textbf{dedicated} asymmetric hardware acceleration ensures results reflect baseline CPU performance relevant to widespread software deployments, with cache hierarchies (L1-L3) and single-threaded execution isolating per-core computational bottlenecks inherent to cryptographic workloads.

\subsection{Benchmarking Software}
The two systems have utilized the Ubuntu Linux distribution, specifically the 24.04.2 LTS version.
On the programming side, Rust was used as the programming language for the benchmarks. The  \href{https://crates.io/crates/openssl}{openssl crate} provided the implementations of RSA and SECP384R1, while the \href{https://crates.io/crates/oqs}{oqs crate} provided ML-KEM via the Rust bindings for the Open Quantum Safe’s liboqs library.

The benchmarks are consolidated into two main sections: Performance and Storage.
The Performance section measures the necessary computational resources required to execute a specific operation, measured in CPU cycles obtained through the use of the  \href{https://crates.io/crates/iai-callgrind}{iai-callgrind} crate. It is essential to note that the CPU cycles are an approximation due to the noise generated by the CPU boost; however, they closely approximate the actual value with a high degree of accuracy.
On the other hand, the Storage section measures the size of the outputs of the ciphers for a given message.
\textbf{The complete benchmarking source code is publicly available on this repo \href{https://github.com/Nichokas/kyber-performance}{nichokas/kyber-performance}.}
Implementations leverage platform-supported hardware acceleration (AES-NI, PMULL, ASIMD) for relevant operations. Asymmetric-specific hardware (e.g., RSA modular exponentiation units) was turned off to ensure algorithmic fairness.

\section{Comparison}
For reproducibility, the version of the source code used for this paper is \href{https://github.com/Nichokas/kyber-performance/tree/6fa6b0ca39312755fc228b50978184df88ca6fb7}{Commit 6fa6b0c}.
\subsection{Speed benchmarks}
\label{sec:speed_benchmarks}
\subsubsection{Key Generation}
Measurement of the required computational resources to create a new public-private keypair.

\begin{figure}[H]
  \centering
  \includegraphics[width=\columnwidth]{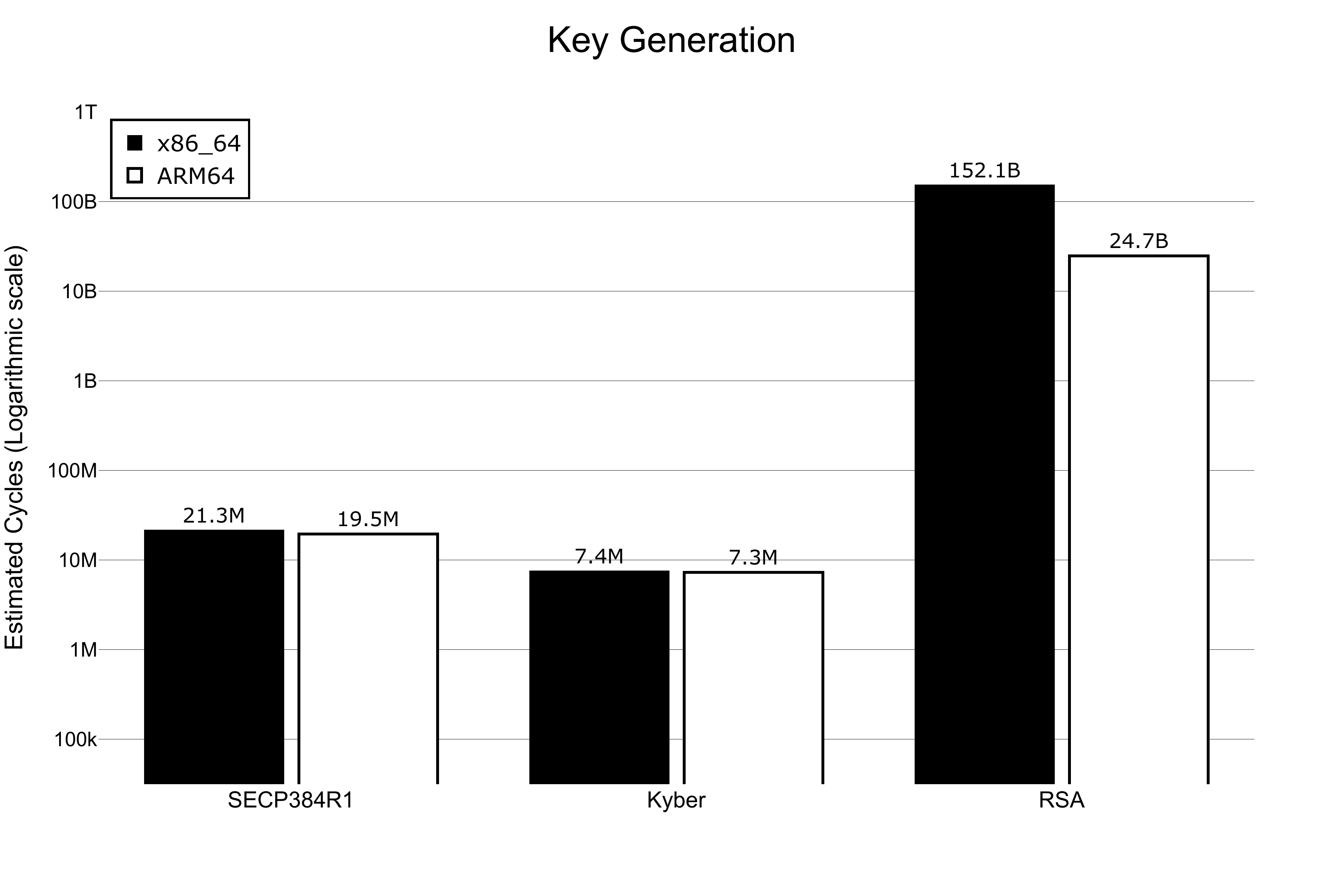}
  \label{fig:speed-plot}
\end{figure}

Kyber shows efficient key generation. On both x86\_64 and ARM64 architectures, Kyber requires the fewest cycles ($\sim$7.4M and $\sim$7.3M, respectively), making it approximately 2.7 to 3 times faster than SECP384R1 ($\sim$21.3M and $\sim$19.5M cycles). In comparison with RSA, the difference is substantial: on the x86\_64 platform Kyber uses roughly 20,500 times fewer cycles, and on ARM64 about 3,400 times fewer. These factors are consistent with the underlying complexity of each algorithm. RSA key generation involves searching for large prime numbers, which is computationally intensive and scales poorly, whereas Kyber’s lattice‑based arithmetic admits more efficient implementations.
Although the ARM64 architecture significantly reduces RSA’s key generation time by a factor of six compared to x86\_64, it is still much slower than both Kyber and ECC.
\subsubsection{Outgoing Shared Secret Derivation}

\begin{figure}[H]
  \centering
  \includegraphics[width=\columnwidth]{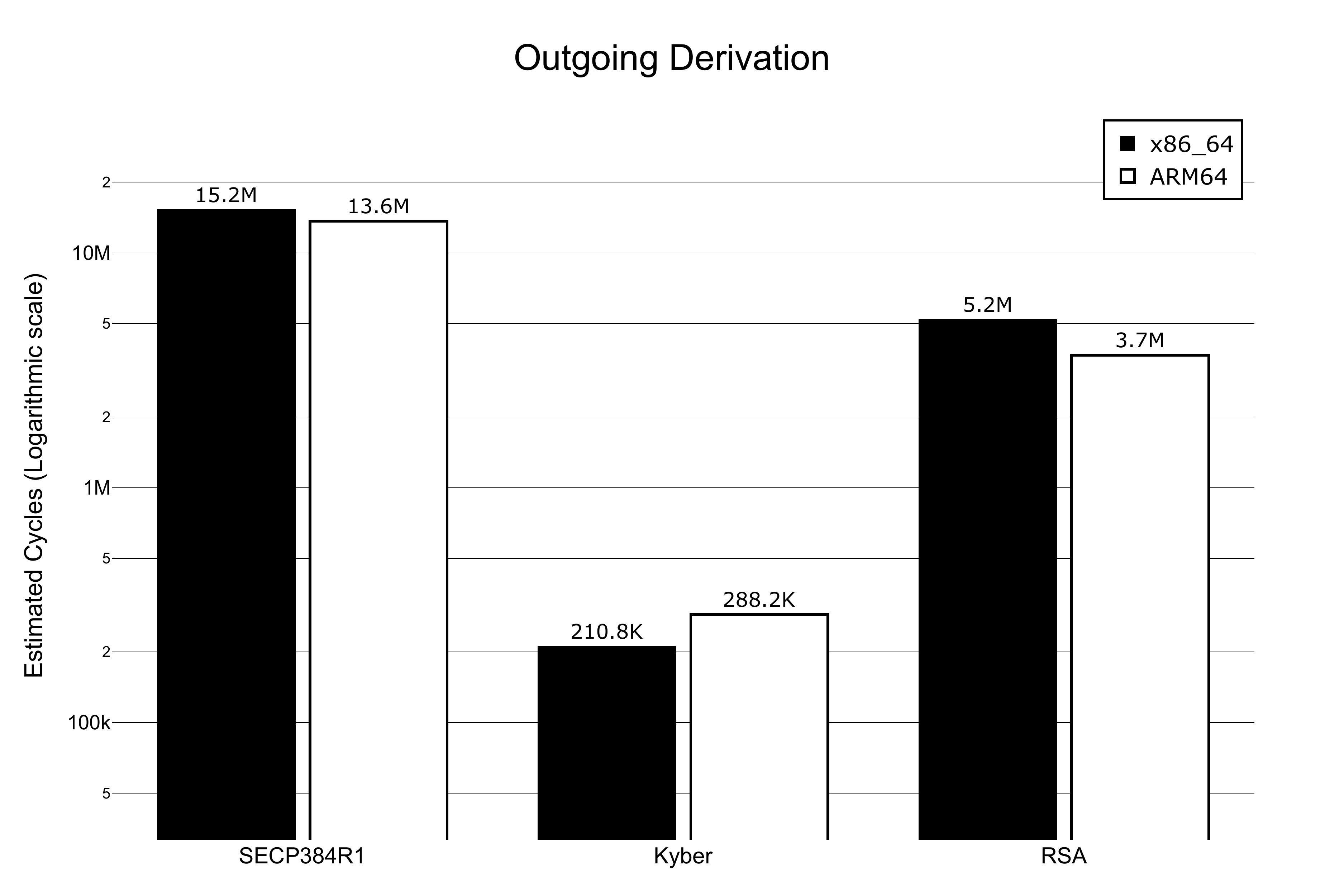}
  \label{fig:outgoing-plot}
\end{figure}
This chart illustrates the performance of deriving a shared secret from the initiator’s (the “outgoing” party’s) perspective. This process typically involves using one’s own private key and the recipient’s public key to establish a mutual secret. The y-axis is on a logarithmic scale to properly visualize the performance differences in estimated CPU cycles.

Kyber is the most efficient algorithm for this operation on both tested platforms. On x86\_64, Kyber’s cost of approximately 210,800 cycles is roughly 25 times lower than RSA ($\sim$5.2 million cycles) and about 72 times lower than SECP384R1 ($\sim$15.2 million cycles). On the ARM64 architecture, Kyber ($\sim$288,200 cycles) maintains its lead, remaining roughly 13 times more efficient than RSA ($\sim$3.7 million cycles) and 47 times more efficient than SECP384R1 ($\sim$13.6 million cycles).

It is important to note that, in RSA, this “public‑key” operation is significantly less costly than the “private‑key” operation shown in the “Incoming Derivation” chart, which is an expected characteristic of the algorithm. Nevertheless, Kyber’s performance remains superior. For SECP384R1, the computational workload is identical for both the incoming and outgoing phases, which explains its consistently high cycle count. Kyber requires fewer than 300,000 cycles for this step, which reduces the latency of cryptographic handshakes relative to both RSA and ECC.

\subsubsection{Incoming Shared Secret Derivation}

\begin{figure}[H]
  \centering
  \includegraphics[width=\columnwidth]{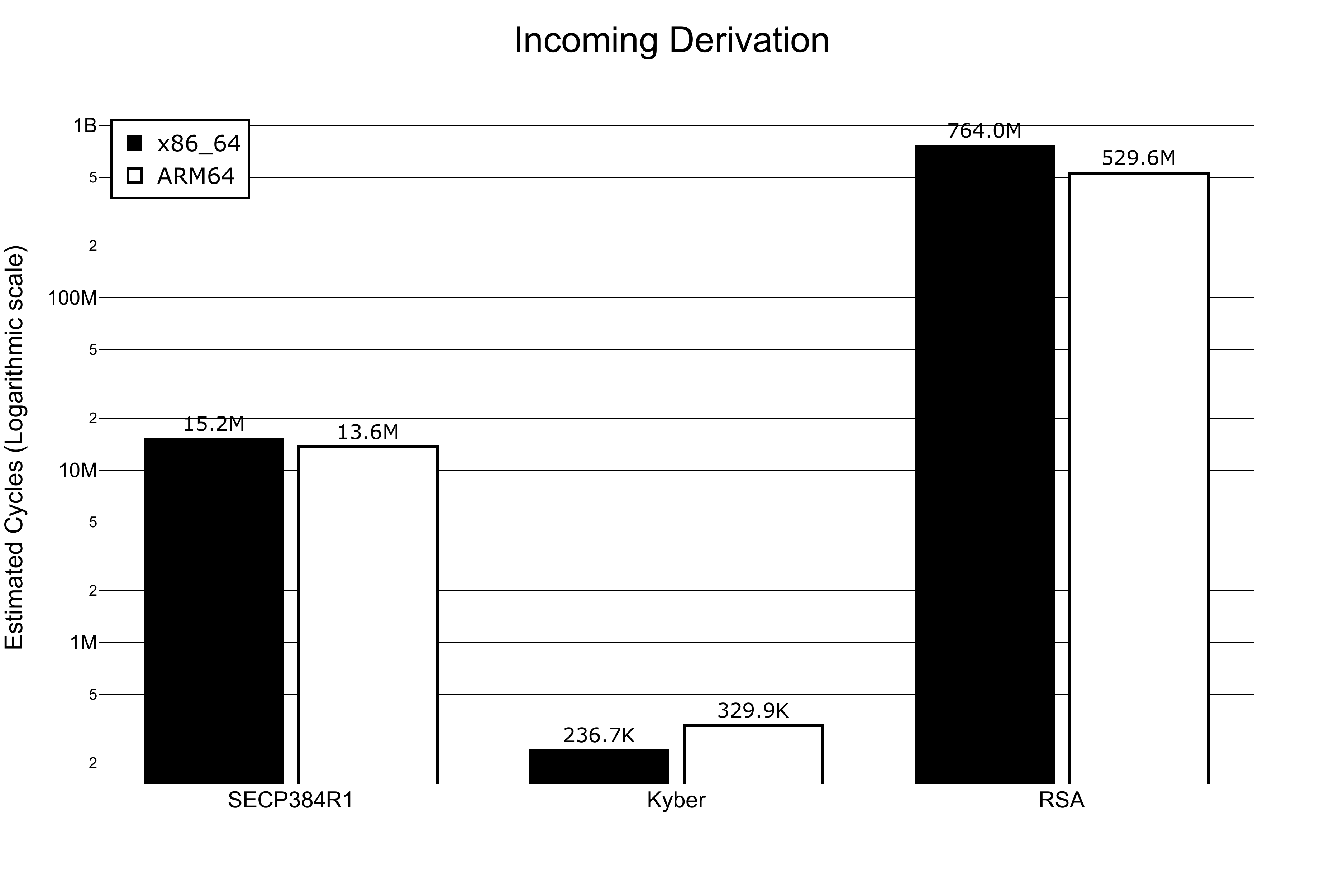}
  \label{fig:incoming-plot}
\end{figure}

On the x86\_64 architecture, Kyber requires about 236,700 cycles for incoming shared secret derivation, whereas SECP384R1 needs 15.2 million cycles and RSA around 764 million cycles. The resulting factors (roughly 64× and 3,200×, respectively) arise from the different mathematical operations underlying each algorithm. A similar trend is observed on the ARM64 platform, where Kyber’s 329,900 cycles outpace SECP384R1 (13.6 million cycles) by about 41× and RSA (529.6 million cycles) by about 1,600×.

For SECP384R1, the derivation procedure is identical for both the outgoing and incoming phases, which explains its consistently high computational cost relative to Kyber in both scenarios. These differences follow from Kyber’s foundation in module lattice‑based arithmetic, which avoids the expensive elliptic curve point multiplication used by SECP384R1 and the even more costly modular exponentiation of RSA.

While the ARM64 architecture reduces the absolute cycle count for classical algorithms, it does not alter the fundamental performance hierarchy.

\subsection{Storage benchmarks}

\begin{figure}[H]
  \centering
  \includegraphics[width=\columnwidth]{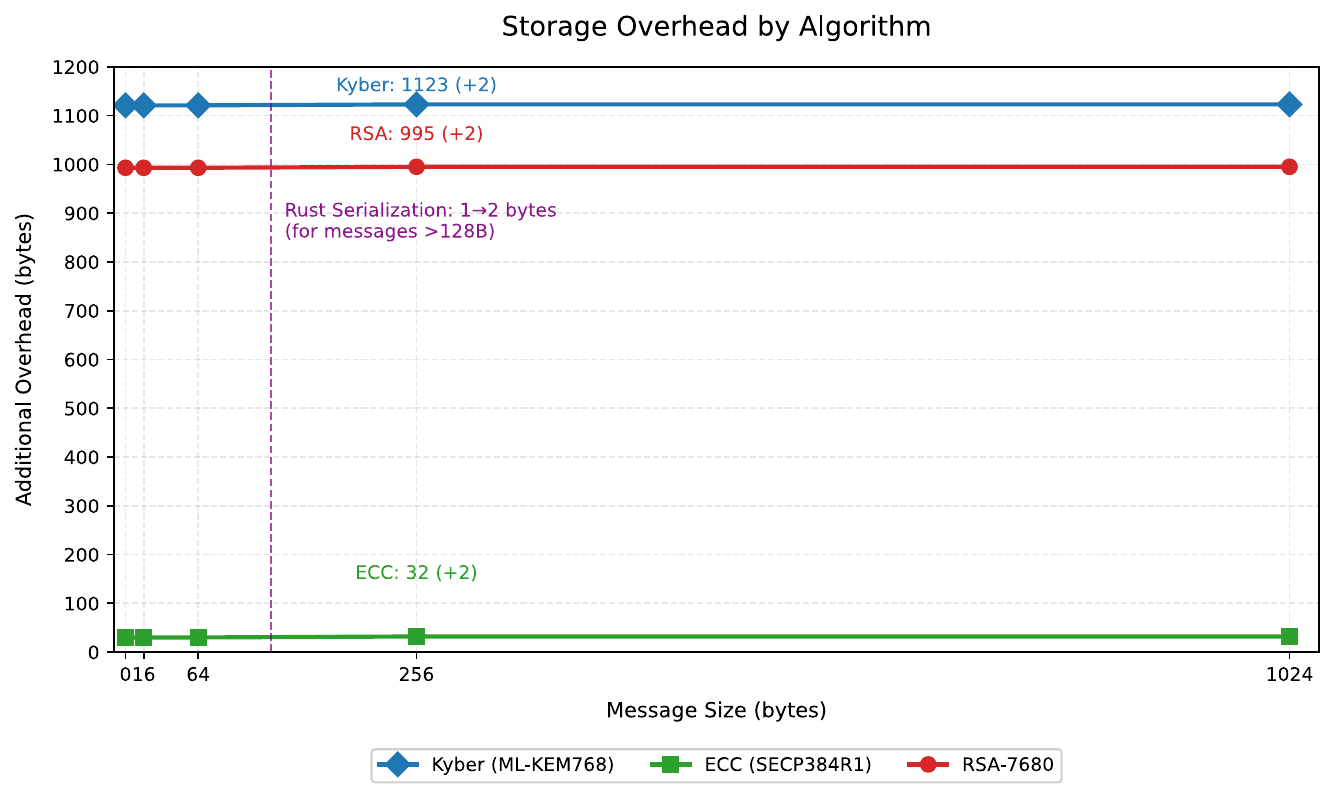}
  \label{fig:storage-plot}
\end{figure}

\begin{table}[H]
  \centering
  \caption{Added overhead (without the Rust bytes)}
  \label{tab:overh}
  \small
  \setlength{\tabcolsep}{3pt} 
  \begin{tabular*}{\linewidth}{@{\extracolsep{\fill}}
      l
      c
      c
      c
    }
    \toprule
     & {nonce} & {Key transport} & {Total added length} \\
    \midrule
    SECP384R1 & 30 &  & 30 \\
    Kyber & 30 & 1091 & 1121 \\
    RSA & 30 & 963 & 993 \\
    \bottomrule
  \end{tabular*}
\end{table}

The added overhead refers to the data that an algorithm adds to the existing data (plaintext length) to encrypt it, and in some cases, to include the key transport. For example, with Kyber, the total is: plaintext message length + Kyber’s overhead + Rust’s serialization bytes.
The storage overhead analysis highlights a trade‑off between post‑quantum security and bandwidth. The Poly1305 authentication nonce adds a constant 30‑byte overhead across all schemes, but the key transport payload varies: SECP384R1 uses elliptic curve key agreement and thus incurs no extra key transport overhead, yielding a total added length of 30 bytes. Kyber and RSA require additional key transport data (1,091 bytes and 963 bytes, respectively), leading to total overheads of 1,121 bytes for Kyber and 993 bytes for RSA. ECC’s minimal overhead makes it the most bandwidth‑efficient option—a consideration for constrained environments such as IoT—but its lack of quantum resistance limits its longevity. Kyber’s roughly $13.3\%$ larger transport payload than RSA reflects the size of its lattice‑based ciphertexts; given its computational advantages and quantum security, this overhead is likely acceptable for many applications.

\section*{Conclusion}
The benchmarking results show that CRYSTALS‑Kyber has significant computational advantages relative to classical schemes across both \texttt{x86\_64} and \texttt{ARM64} architectures. These advantages arise from Kyber’s use of lattice‑based polynomial arithmetic, which benefits from vectorization instructions such as \textbf{AVX2}. In our tests, Kyber’s key generation required roughly \textbf{2.7--3$\times$} fewer cycles than ECC and many orders of magnitude fewer than RSA, while shared secret derivation also exhibited multi‑order‑of‑magnitude speed differences. These results are broadly consistent across architectures; the \texttt{ARM64} platform reduces the absolute cost of classical schemes but does not change the relative ordering. Kyber’s hybrid ciphertexts have a larger payload—about 1,121 bytes compared to RSA’s 993 bytes and ECC’s minimal 30 bytes—which reflects the trade‑off between post‑quantum security and bandwidth efficiency. Overall, our experiments indicate that Kyber can deliver NIST‑standardized quantum resistance using current commodity hardware without incurring prohibitive computational costs.

These findings suggest that, with existing CPU support for instructions like \textbf{AVX2} and \textbf{AES-NI}, lattice‑based cryptography is practical today. Future development of specialized hardware accelerators for lattice operations could further reduce latency and power consumption, expanding the range of applications—including resource‑constrained environments such as IoT devices—where post‑quantum cryptography may be deployed.

\section*{Acknowledgements}
I would like to extend my sincere gratitude to Elías F. Combarro, Full Professor in the Department of Computer Science at the University of Oviedo (Spain), for his invaluable mentorship on the intricacies of academic writing and for the constructive feedback he provided during the development of this manuscript. His support was instrumental in shaping this work.

I am also grateful to Fernando Rodriguez Merino\orcidlink{0000-0002-3991-2563} (Department of Theoretical, Atomic and Optical Physics, University of Valladolid, Spain) for putting me in contact with Prof. Combarro and for occasional discussions and feedback over the course of this work.

\bibliographystyle{alpha}
\bibliography{biblio}

\twocolumn[
\section{Raw Obtained Data}
\label{sec:raw_data}
\centering
\captionsetup{type=table}
\captionof{table}{Consolidated Cryptographic Performance Metrics (x86\_64 \& ARM64)}
\perfstyle
\small
\setlength{\tabcolsep}{4pt}
\begin{tabular}{lllrrrrr}
\toprule
\textbf{Algorithm} & \textbf{Operation} & \textbf{Arch} & 
\textbf{Instr} & \textbf{L1 Hits} & \textbf{L2 Hits} & 
\textbf{RAM Hits} & \textbf{Est. Cycles} \\
\midrule
SECP384R1 & Incoming Secret & x86\_64 & 11658237 & 15117387 & 4505 & 1912 & 15206832 \\
SECP384R1 & Incoming Secret & ARM64   & 10231635 & 13557483 & 4534 & 1859 & 13645218 \\
SECP384R1 & Outgoing Secret & x86\_64 & 11658237 & 15117387 & 4505 & 1912 & 15206832 \\
SECP384R1 & Outgoing Secret & ARM64   & 10231635 & 13557483 & 4534 & 1859 & 13645218 \\
SECP384R1 & Key Generation  & x86\_64 & 15975363 & 20939189 & 12050 & 9278 & 21324169 \\
SECP384R1 & Key Generation  & ARM64   & 14259264 & 19082149 & 12224 & 9094 & 19461559 \\

\midrule
Kyber & Incoming Secret & x86\_64 & 177064   & 212675   & 2211 & 370  & 236680   \\
Kyber & Incoming Secret & ARM64   & 268713   & 319023   & 413  & 251  & 329873   \\
Kyber & Outgoing Secret & x86\_64 & 152513   & 184443   & 2332 & 419  & 210768   \\
Kyber & Outgoing Secret & ARM64   & 229406   & 272873   & 833  & 318  & 288168   \\
Kyber & Key Generation  & x86\_64 & 5266214  & 7050285  & 10400 & 9543 & 7436290  \\
Kyber & Key Generation  & ARM64   & 5082726  & 6893944  & 10393 & 8955 & 7259334  \\

\midrule
RSA & Incoming Secret   & x86\_64 & 609884957   & 763988457  & 4563    & 665    & 764034547   \\
RSA & Incoming Secret   & ARM64   & 459230402   & 528209769  & 273797  & 494    & 529596044   \\
RSA & Outgoing Secret   & x86\_64 & 4229026     & 5169943    & 1523    & 335    & 5189283     \\
RSA & Outgoing Secret   & ARM64   & 3124179     & 3652723    & 1507    & 266    & 3669568     \\
RSA & Key Generation    & x86\_64 & 121603553478 & 152134665100 & 942996  & 20601  & 152140101115 \\
RSA & Key Generation    & ARM64   & 21637354719  & 24672791454  & 10775986 & 20933  & 24727404039  \\
\bottomrule
\end{tabular}
\vspace{1em}
]

\end{document}